# Raman scattering of perovskite SmScO$_3$ and NdScO$_3$ single crystals

O. Chaix-Pluchery[1] and J. Kreisel

*Laboratoire des Matériaux et du Génie Physique,*
*CNRS UMR 5628 - Grenoble Institute of Technology - Minatec, 3, parvis Louis Néel, 38016 Grenoble, France*

**Abstract**

We report an investigation of perovskite-type SmScO$_3$ and NdScO$_3$ single crystals by Raman scattering in various scattering configurations and at different wavelengths. The reported Raman spectra, together with the phonon mode assignment, set the basis for the use of Raman scattering for the structural investigation of RE-scandates. Further to the phonon signature, a fluorescence signal is observed for both scandates and is particularly intense for NdScO$_3$ when using a 488 or 514 nm excitation line. A comparison of Raman spectra of *RE*-scandates with literature Raman data on orthorhombic perovskites shows that the frequency of particular modes scales with the orthorhombic distortion in terms of the rotation (or tilt) angle of the ScO$_6$ octahedra.

[1] to whom correspondence should be addressed (*E-mail: Odette.Chaix@grenoble-inp.fr*)

## I. INTRODUCTION

The understanding of functional $AB$O$_3$ perovskite-type oxides is a very active research area with relevance to both fundamental- and application-related issues. A particular aspect of perovskites is its capacity to incorporate almost every element of the periodic table into its structure, leading to a multitude of observed structural distortions[1] which, in turn, conditions an extraordinary richness of physical properties within the family of perovskites.

Among perovskites, orthorhombic rare earth (*RE*) scandates *RE*ScO$_3$ receive since recently an increasing interest, due to their potential in mainly three different fields : (*i*) *RE*-scandates are regarded as a promising high-κ dielectric material candidate for the replacement of SiO$_2$ in silicon MOSFET [2-8]; (*ii*) *RE*-scandate single crystals are considered to be among the best available substrates for the epitaxial growth of high-quality perovskite-type thin films, namely aiming at strain engineering of ferroelectric and multiferroic properties[9-14]; (*iii*) Scandates embedded in thin film perovskite heterostructures (i.e. SrTiO$_3$/DyScO$_3$ or BaTiO$_3$/GdScO$_3$) are currently investigated for applications in the terahertz range[15] or again as high-κ materials [8].

Advances in all these three fields depend on appropriate techniques for a detailed structural characterisation. In this study we focus on phonon Raman spectroscopy, which is known to be a versatile technique for the investigation of subtle structural distortions in perovskites [16-22] and, more generally, for the understanding of thin film characteristics such as strain effects [23-25], texture [17], X-ray amorphous phases [19,26], heterostructure-related features [20,27,28]. Contrary to other orthorhombic perovskites (i.e. ortho-ferrites, -chromites, -manganites, or –nickelates) for which the Raman signature is rather well known and understood, the understanding of ortho-scandates remains limited to few Raman studies on DyScO$_3$ and GdScO$_3$. [6,29,30]

We present an experimental Raman scattering study of SmScO$_3$ and NdScO$_3$ single crystals with the aim (*i*) to provide reference data for future Raman investigations of bulk and thin film *RE*-scandates and (*ii*) to extend an earlier Raman study[30] of DyScO$_3$ and GdScO$_3$ in order to get a better understanding of the vibrational signature of *RE*-scandates via the systematic comparison to other orthorhombic perovskites, namely the well-investigated[31] family of ortho-manganites.

## II. EXPERIMENTAL AND SAMPLE CHARACTERISATION

SmScO$_3$ and NdScO$_3$ single crystal platelets, 5*5*0.5 mm$^3$ of size, have been investigated by Raman spectroscopy. The crystals provided from Crystec have been grown by the conventional Czochralski technique as described in ref [32]. The starting materials for the crystal growth processes, Sm$_2$O$_3$, Nd$_2$O$_3$ and Sc$_2$O$_3$, were of 99.99% purity. The crystal platelet orientation is shown in Fig. 1: their straight edges are parallel to the pseudo-cubic cell axes, i.e. to the [010] (=y) and [-101] (=x') orthorhombic axes, as confirmed by polarization properties of the Raman spectra.

Raman spectra were recorded with a Jobin-Yvon/Horiba LabRam spectrometer equipped with a $N_2$-cooled CCD detector with using three different wavelengths as exciting lines (488.0, 514.5 and 632.8 nm). Experiments were conducted in micro-Raman mode at room temperature, the used configurations of polarization are identical to those of our earlier work on Dy- and GdScO$_3$ [30].

## III. RESULTS AND DISCUSSION

*RE*-scandates crystallize in an orthorhombically distorted perovskite structure with space group *Pnma*. With respect to the ideal cubic *Pm*3*m* perovskite structure this orthorhombic structure is obtained by an anti-phase tilt of the adjacent ScO$_6$ octahedra ($a^-b^+a^-$ in Glazer's notation [33]). The orthorhombic *Pnma* structure is commonly observed in perovskites and the four formula units per unit cell give rise to 24 Raman-active modes [34]: 7 $A_g$ + 5 $B_{1g}$ + 7 $B_{2g}$ + 5 $B_{3g}$. For comparison, note that Raman spectra of several *Pnma* orthorhombic perovskites have been reported in literature: e.g orthoferrites *RE*FeO$_3$ [35,36], orthochromites *RE*CrO$_3$ [37], orthomanganites *RE*MnO$_3$ [31,34], nickelates *RE*NiO$_3$ [18,38] or at high pressure the model multiferroic BiFeO$_3$ [39].

The depolarized spectra of SmScO$_3$ and NdScO$_3$ collected on the crystal surface using the 632.8 nm exciting wavelength are shown in Fig. 2; for comparison earlier published Raman spectra[30] of DyScO$_3$ and GdScO$_3$ are also shown. Qualitatively, the overall spectral signature of the four scandates is very similar, in agreement with the fact that they adopt the same crystal structure with an orthorhombic distortion which differs only little due to changes in $RE^{3+}$ ionic radii within 10%.[40] The difference in relative intensities from Dy- and GdScO$_3$, on the one hand, to Sm- and NdScO$_3$ on the other hand is due to the orthogonal orientations of both series of crystals (surface normal parallel to the y axis for the two first ones, perpendicular to the y axis for the two last ones). In the following, we will assign the symmetry of the different modes in SmScO$_3$ and NdScO$_3$ from their polarized Raman spectra and compare the Raman line positions in the four *RE*-scandate spectra taking into account the $RE^{3+}$ ionic radius influence. Motivated by the effects observed earlier[30] for DyScO$_3$ and GdScO$_3$, we will examine the effect of the laser wavelength on the spectra of SmScO$_3$ and NdScO$_3$.

### A. Symmetry assignment of Raman modes in SmScO$_3$ and NdScO$_3$

Polarized Raman spectra of SmScO$_3$ and NdScO$_3$ are reported in Fig. 3a and 3b, respectively. As illustrated in the figures, three distinct polarization configurations (Porto's notation) have been used to identify the different symmetry modes. The parallel *z'(yy)z'* and *z'(x'x')z'* polarization configurations allow to observe the $A_g$ and $A_g$+$B_{2g}$ modes, respectively: as shown in Fig. 3, $A_g$ modes occur at 110, 130, 233, 312, 400, 442 and 494 cm$^{-1}$ in SmScO$_3$ and at 111, 131, 227, 302, 380, 433 and 492 cm$^{-1}$ in NdScO$_3$. The comparison of spectra in both polarization configurations led us to point $B_{2g}$ modes at 115, 155, 170,

279, 343, 449, 523 cm$^{-1}$ in SmScO$_3$, at 117, 154, 167, 259, 340, 434, 513 cm$^{-1}$ in NdScO$_3$. The crossed $z'(x'y)z'$ polarization configuration then allows measurement of the modes assigned as $B_{1g}$ or $B_{3g}$ but this configuration does not allow their experimental distinction in terms of symmetry. On the other hand, the comparison of modes in SmScO$_3$ and NdScO$_3$ with previously assigned modes in DyScO$_3$ and GdScO$_3$ leads to the assignment of $B_{1g}$ and $B_{3g}$ modes presented in Table 1. A good overall correspondence between modes of the four $RE$-scandates is observed, although the first low wavenumber $B_{1g}$ and $B_{3g}$ modes are not discriminated and the highest wavenumber $B_{2g}$ mode is only observed in the DyScO$_3$ spectrum. An additional mode of apparent $B_{2g}$ symmetry occurs at 170 and 167 cm$^{-1}$ in SmScO$_3$ and NdScO$_3$ spectra, respectively, as already observed for DyScO$_3$ and GdScO$_3$ at 176 and 174 cm$^{-1}$, respectively. Its origin and assignment remain to be elucidated.

## B. Influence of the rare earth on *RE*-scandate Raman spectra

In the past, the empirical comparison of materials with the same type of polyhedra (i.e. [41-43]) or the same type of structural distortion (i.e. [31,34,44]) has been a rich source of understanding Raman signatures in oxide materials. For the specific case of orthorhombic perovskites it is useful to remind the systematic Raman work by Iliev *et al.*[31] of orthorhombic $RE$MnO$_3$ manganites which crystallize in the same *Pnma* symmetry as $RE$-scandates. Among other results, it has been shown that the frequency of some $A_g$ modes in $RE$MnO$_3$ manganites scales with the rotation angle of the MnO$_6$ octahedral tilting[31]. Just as in $RE$-manganites, the orthorhombic distortion in $RE$-orthoscandates depends mainly on the ionic radii $r_{RE(3+)}$. As a basis of the following discussion of the Raman signature of scandates, Table 2 presents a summary of $r_{RE(3+)}$, the octahedral tilt angle, the lattice parameters but also the $RE$ masses $m_{RE(3+)}$ for Dy-, Gd-, Sm- and NdScO$_3$. The dependence of the A$_g$ and B$_{2g}$ mode wavenumbers on $r_{RE(3+)}$ in Dy-, Gd-, Sm- and NdScO$_3$ is shown in Fig. 4. The modes can be divided into two groups: (1) Four modes below 200 cm$^{-1}$, which change only little in wavenumber as a function of $r_{RE(3+)}$. (2) Modes above 200 cm$^{-1}$, all of which shift significantly to lower wavenumber with increasing $r_{RE(3+)}$.

For comparing different materials of the same symmetry, it is useful to remind the approximation of the harmonic oscillator $\omega = (k/\mu)^{1/2}$ (k: force constant, µ: reduced mass) which allows estimating the effect of changing at the same time $r_{RE(3+)}$ and $m_{RE(3+)}$. It is readily seen that heavier atoms vibrate at lower frequencies, thus the modes of the first group below 200 cm$^{-1}$ are expected to be dominated by vibration involving the $RE$ atom, which is the heaviest atom in the $RE$ScO$_3$ structure. On the other hand, an increasing ionic radius leads to increased bond lengths (and lattice parameters) and thus in turn to lower force constants $k$ and thus decreasing frequencies.

Based on this, it becomes clear that the rather $RE$-independent behavior of the modes below 200 cm$^{-1}$ (group 1) is a result of mutual neutralizing effects of increasing mass and decreasing ionic radii. On the other hand, above 200 cm$^{-1}$, all modes are significantly and linearly shifted towards low wavenumbers with increasing $r_{RE(3+)}$. The evolution of all these modes is thus rather dominated by increasing bond lengths than by mass effects, in agreement with the fact that the lighter atoms of the structure (scandium and oxygen) do not change. It is interesting to note that three bands show a particularly pronounced shift in position with increasing $r_{RE(3+)}$: $B_{2g}(3)$, $A_g(5)$ and $B_{2g}(5)$. This observation can be understood by considering that some modes capture specifically changes in the octahedral tilts, i.e. the structural instability (order parameter) of *RE*-scandates with respect to ideal cubic perovskite structure (increasing $r_{RE(3+)}$ conditions a reduction of $ScO_6$ octahedral tilting[45], see Table 2 for explicit values). In this light it is interesting to remind the work by Iliev *et al.* who proposed that specific rotational $A_g$ modes scale in *RE*MnO$_3$ manganites with the rotation angle of the MnO$_6$ octahedral tilting [31]. Applying the same reasoning to our work we find that the $A_g(3)$ and $A_g(5)$ modes in the scandates (labeled $A_g(2)$ and $A_g(4)$, respectively in Iliev's work[31]) are rotational $A_g$ modes and thus naturally probe changes in the octahedral tilt angle. Figure 5 presents the wavenumber of the two $A_g$ rotational modes as a function of the corresponding octahedra tilt angle for *RE*-scandates together with earlier published data for manganites from ref. [31] A first inspection of the figure shows that the positions of the $A_g(3)$ and $A_g(5)$ modes in the scandates show qualitatively the same tendency as the manganite modes and lie surprisingly close to Iliev's proposed slope of $\omega$ = f(tilt angle) with $\approx$ 23.5 cm$^{-1}$/deg. However, it can also be seen that a proper and more appropriate slope of $\approx$ 20 cm$^{-1}$/deg can be defined for scandates. This observation in scandates, thus in another family than manganite-type orthorhombic perovskites, adds further support that Iliev's proposition[31] of a general relationship between the angle of octahedral tilting and the frequency of the rotational soft modes is viable in a vast number of perovskites. On the other hand, our study suggests that the actual slope depends on the transition metal on the *B*-site.

## C. Influence of exciting wavelength on SmScO$_3$ and NdScO$_3$ Raman spectra

Raman spectra of Sm- and NdScO$_3$ collected by using three different exciting laser lines (488, 514.5 and 632.8 nm) are reported in Fig. 6. In the case of SmScO$_3$ we observe a similar spectrum whatever the exciting wavelength (Fig. 6a) whereas fluorescence lines occur for NdScO$_3$ (Fig. 6b): The fluorescence signal is superimposed to the NdScO$_3$ Raman spectrum when using the 514.5 nm exciting line, being very high in the 600-1100 cm$^{-1}$ range, while it is much weaker when using the 632.8 nm line (above 600 cm$^{-1}$) or the 488 nm line (above 1500 cm$^{-1}$).

For a further characterization of the fluorescence in the *RE*-scandates, spectra have then been measured for NdScO$_3$ and SmScO$_3$ from 490 to 870 nm using the three exciting lines. As seen in Fig. 7, several groups of fluorescence lines are observed in both spectra: very intense lines, in comparison to Raman lines (labeled $R_\lambda$), are observed for NdScO$_3$ as well as a very strong signal from 870 nm with any of the three exciting lines (Fig. 7b). The intense lines are grouped together at wavelengths close to 600, 675 and 820 nm; weak additional lines are also observed in the 520-550 and 740-780 nm ranges. The strong emission above 870 nm is related to the typical emission of Nd$^{3+}$ in the near infrared range (850-955 and 1045-1080 nm [46,47]. All other lines are indexed in the absorption spectra of Nd$^{3+}$ doped films of ref. [47]

Fluorescence spectra measured for SmScO$_3$ (Fig. 6a) are composed of four groups of lines near 565, 610, 655 and 715 nm when excited with the 514.5 or 488 nm line, of variable intensities depending on the exciting wavelength but much less intense than NdScO$_3$ fluorescence lines. All correspond to typical emission bands of Sm$^{3+}$ as observed in Sm$^{3+}$-doped Al$_2$O$_3$ films excited at 404 nm [48] or in solutions of perchlorate salts of Sm$^{3+}$ excited at 561 nm [49]. Only a weak and very large band is visible between 700 and 850 nm when using the 632.8 nm exciting line. Its origin remains unknown but this broad band could be related to impurities contained in the starting binary oxides; we note that a similar but more intense band was observed in the GdScO$_3$ fluorescence spectrum with the same excitation line [30].

## IV. CONCLUSION

We have presented a Raman scattering investigation of SmScO$_3$ and NdScO$_3$ single crystals using three different excitations: 488, 514.5 and 632.8 nm. A symmetry assignment of the 23 observed modes has been proposed on the basis of the analysis of various configurations of polarisation and the comparison with our earlier published data on other scandates[30]. The presented experimental data set the basis for the use of Raman scattering as a structural probe in scandates, e.g. for the investigation of strain (via phonon shifts) or texture (via polarised measurements) in thin films or potential phase transitions under temperature or pressure. Similarly to the case of GdScO$_3$ [30], the Raman analysis of NdScO$_3$ should be performed with a 632.8 nm excitation to avoid that a fluorescence signal masks the Raman modes. A further characterisation of the fluorescence in SmScO$_3$ and NdScO$_3$ allowed assigning the fluorescence lines in the 490 to 870 nm range to typical emission bands of the rare earths Sm$^{3+}$ and Nd$^{3+}$, respectively.

A comparison of Raman spectra both within the Dy-, Gd-, Sm- and NdScO$_3$ series and with published Raman data[31] on ortho-manganites *RE*MnO$_3$ shows that the frequency of particular modes, e.g. the rotational $A_g$ modes, scales with the orthorhombic distortion condition as expressed by the rotation (or tilt) angle of the ScO$_6$ octahedra.

TABLE 1: Phonon wavenumbers experimentally obtained from polarized Raman spectra of $RE$ScO$_3$ single crystals ($RE$ = Dy, Gd, Sm, Nd).

| Symmetry mode | DyScO$_3$ $\omega_{exp}$ (cm$^{-1}$) | GdScO$_3$ $\omega_{exp}$ (cm$^{-1}$) | SmScO$_3$ $\omega_{exp}$ (cm$^{-1}$) | NdScO$_3$ $\omega_{exp}$ (cm$^{-1}$) |
|---|---|---|---|---|
| $A_g(1)$ | 111 | 113 | 110 | 111 |
| $A_g(2)$ | 131 | 131 | 130 | 131 |
| $A_g(3)$ | 254 | 248 | 233 | 227 |
| $A_g(4)$ | 327 | 321 | 312 | 302 |
| $A_g(5)$ | 434 | 418 | 400 | 380 |
| $A_g(6)$ | 459 | 452 | 442 | 433 |
| $A_g(7)$ | 509 | 501 | 494 | 492 |
| $B_{2g}(1)$ | 112 | 115 | 115 | 117 |
| $B_{2g}(2)$ | 156 | 155 | 155 | 154 |
| ? | *176* | *174* | *170* | *167* |
| $B_{2g}(3)$ | 309 | 298 | 279 | 259 |
| $B_{2g}(4)$ | 355 | 351 | 343 | 340 |
| $B_{2g}(5)$ | 475 | 463 | 449 | 434 |
| $B_{2g}(6)$ | 542 | 532 | 523 | 513 |
| $B_{2g}(7)$ | 662 | … | … | ... |
| $B_{3g}(1)$ | 104 ? | 110 ? | (101?) | (107?) |
| $B_{3g}(2)$ | 301 | 300 | 296 | 297 |
| $B_{3g}(3)$ | 457 | 450 | 440 | 432 |
| $B_{3g}(4)$ | 477 | 481 | 472 | 470 |
| $B_{3g}(5)$ | 666 | 669 | 665 | 667 |
| $B_{1g}(1)$ | 104 ? | 110 ? | 115? | 122? |
| $B_{1g}(2)$ | 226 | 223 | 219 | 205 |
| $B_{1g}(3)$ | 381 | 373 | 366 | 359 |
| $B_{1g}(4)$ | 492 | 490 | 481 | 483 |
| $B_{1g}(5)$ | (582) | (585) | 615 | 618 |

TABLE 2: Structural characteristics of the *RE*-scandates: lattice parameters[a], $RE^{3+}$ ionic radii[b] ($r_{RE(3+)}$ values given in an eight-fold environment), octahedral tilt angles[a] ($\Psi_{[101]}$ and $\Psi_{[010]}$) and *RE* atomic masses ($m_{RE(3+)}$).

|  | *a* (Å) | *b* (Å) | *c* (Å) | $r_{RE(3+)}$ (Å) | $\Psi_{[101]}$ (°) | $\Psi_{[010]}$ (°) | $m_{RE(3+)}$ |
|---|---|---|---|---|---|---|---|
| **DyScO₃** | 5.717(2) | 7.901(2) | 5.443(2) | 1.027 | 21.00 | 13.23 | 162.50 |
| **GdScO₃** | 5.745(1) | 7.929(2) | 5.481(1) | 1.053 | 20.03 | 12.83 | 157.25 |
| **SmScO₃** | 5.758(2) | 7.975(2) | 5.531(1) | 1.079 | 19.28 | 12.42 | 150.36 |
| **NdScO₃** | 5.777(2) | 8.005(2) | 5.577(1) | 1.109 | 18.07 | 12.04 | 144.24 |

[a]Reference 45.

[b]Reference 40.

# Figure captions

FIG. 1: Schematic representation of the orientation of the $SmScO_3$ and $NdScO_3$ crystal platelets in the orthorhombic unit cell.

FIG. 2: Comparison between depolarized Raman spectra of $REScO_3$ single crystals ($RE$ = Dy, Gd, Sm, Nd) collected with using the 632.8 nm He-Ne laser line. The propagation direction of the incoming and collected photons is along the $y$-axis of the crystal platelets for Dy- and $GdScO_3$, along the $z'$-axis for Sm- and $NdScO_3$. Arrows denote $A_g(3)$ and $A_g(5)$ modes in the four spectra.

FIG. 3: Polarized Raman spectra of $SmScO_3$ (a) and $NdScO_3$ (b) collected in various polarization configurations with using the 632.8 nm He-Ne laser line. The positions of phonon modes allowed in each configuration are given in bold [$A_g$ in $z'(yy)z'$, $A_g + B_{2g}$ in $z'(x'x')z'$, $B_{1g} + B_{3g}$ in $z'(x'y)z'$].

FIG. 4: Dependence of the $A_g$ and $B_{2g}$ mode wavenumbers on the $RE$ ionic radii ($r_{RE(3+)}$) in $REScO_3$ ($RE$ = Dy, Gd, Sm, Nd).

FIG. 5: $A_g(3)$ [open circles] and $A_g(5)$ [full circles] phonon wavenumbers as a function of octahedra tilt angles in $REScO_3$, in comparison with other $RE$ perovskites ($REMnO_3$ data and vibrational pattern taken from Iliev et al. [31], $LaAlO_3$ from Scott [50] and $LaNiO_3$ from [51]).

FIG. 6: Dependence of the depolarized Raman spectra of $SmScO_3$ (a) and $NdScO_3$ (b) on the exciting laser line ($\lambda$ = 632.8, 514.5 and 488 nm).

FIG. 7: $SmScO_3$ (a) and $NdScO_3$ (b) spectra of Fig. 6 presented in a wavelength scale (nm). The lines under the symbol "$R_\lambda$" correspond to the Raman part of the spectra collected with using $\lambda$ = 488, 514.5 and 632.8 nm as exciting lines.

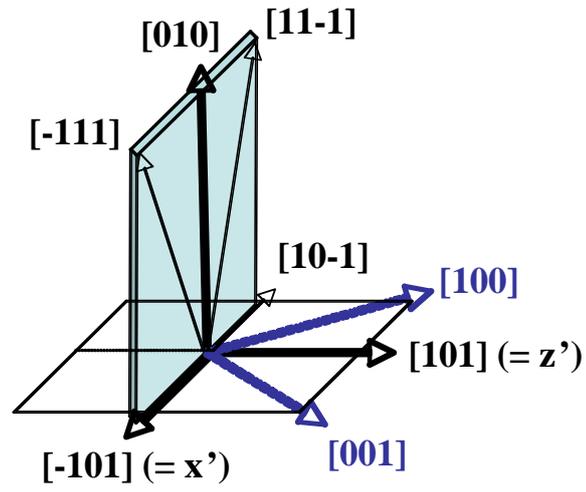

Figure 1

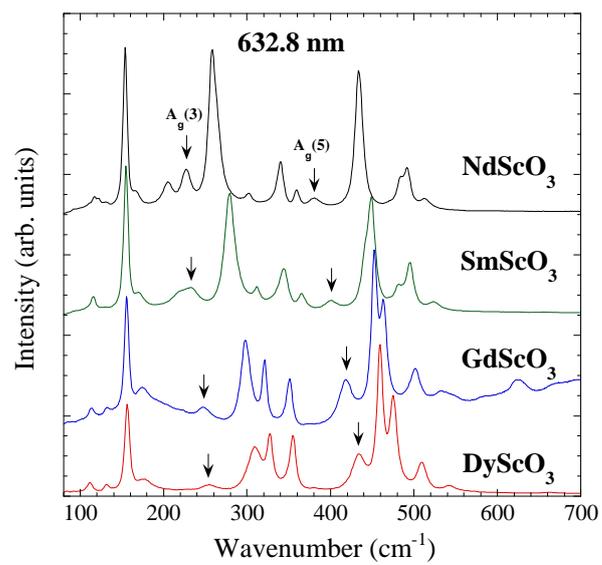

Figure 2

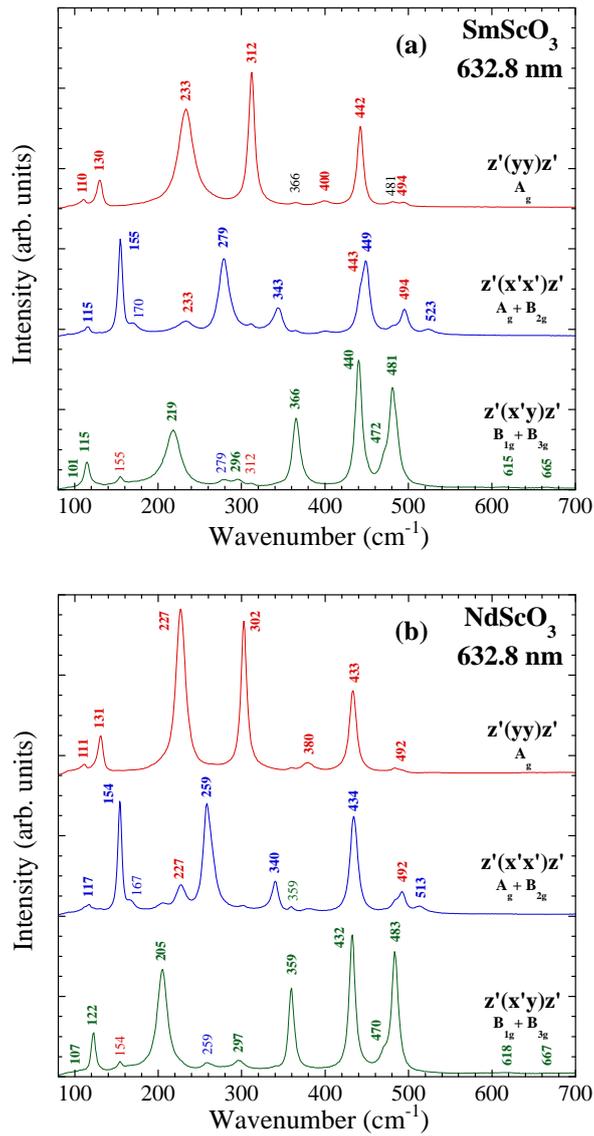

Figure 3

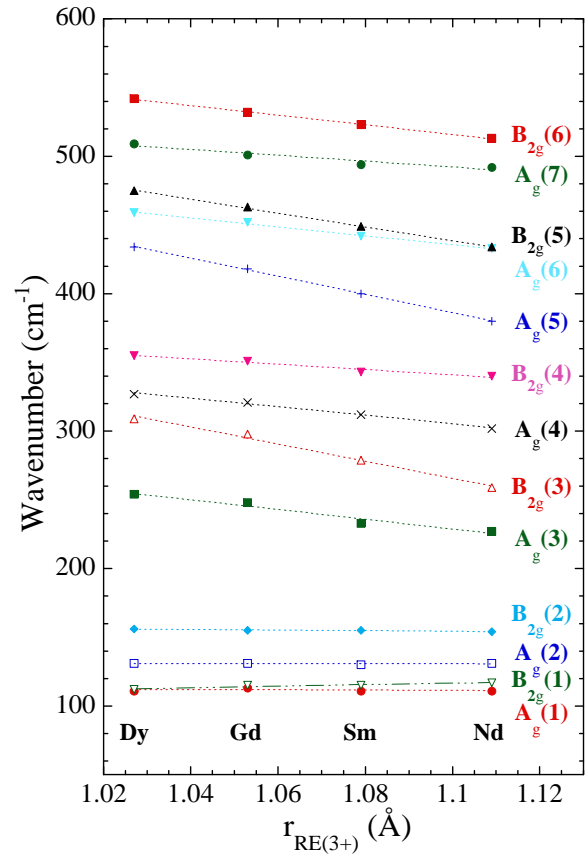

Figure 4

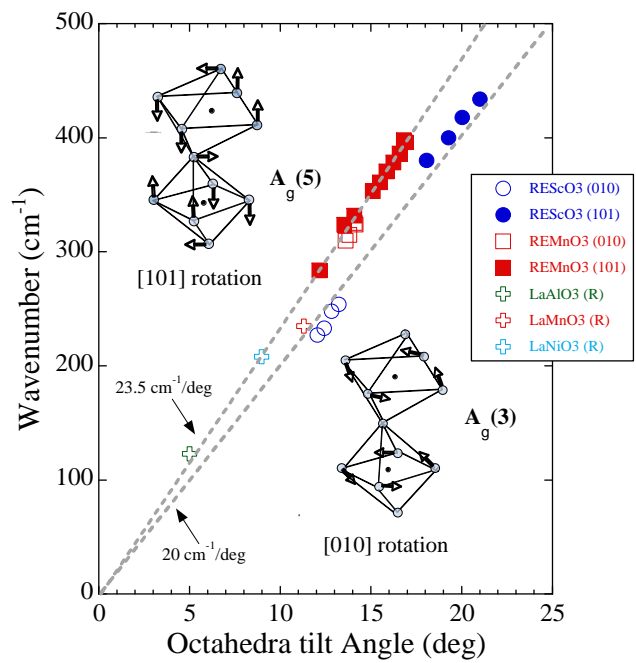

Figure 5

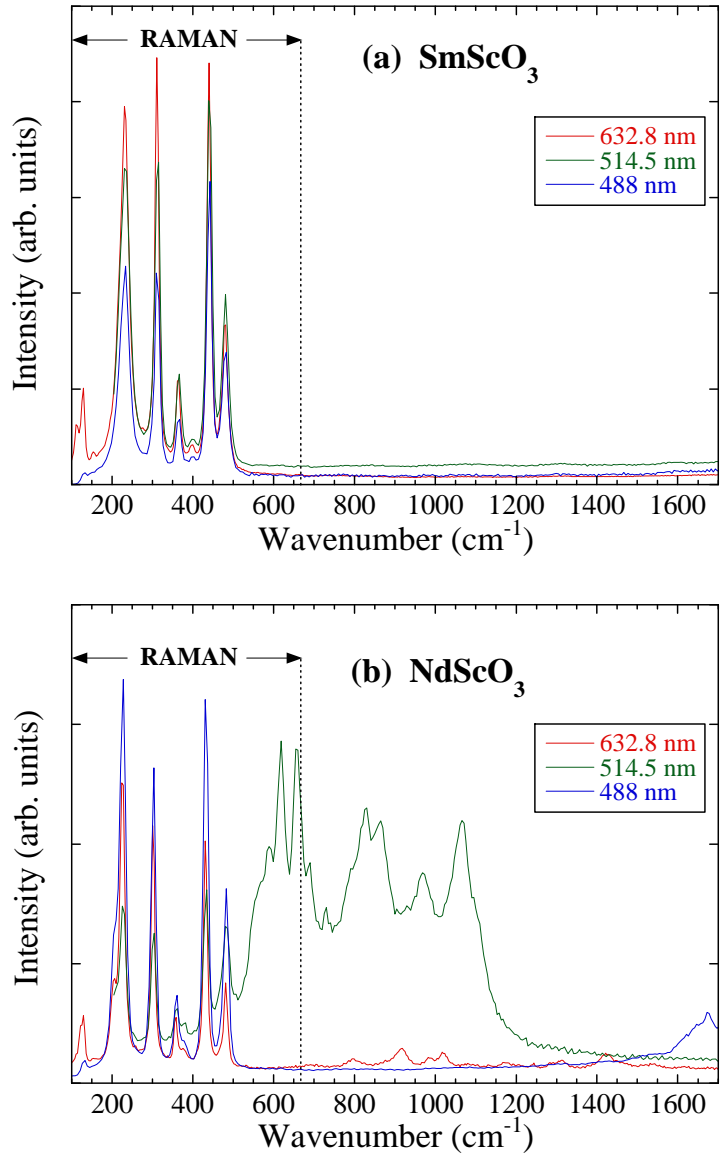

Figure 6

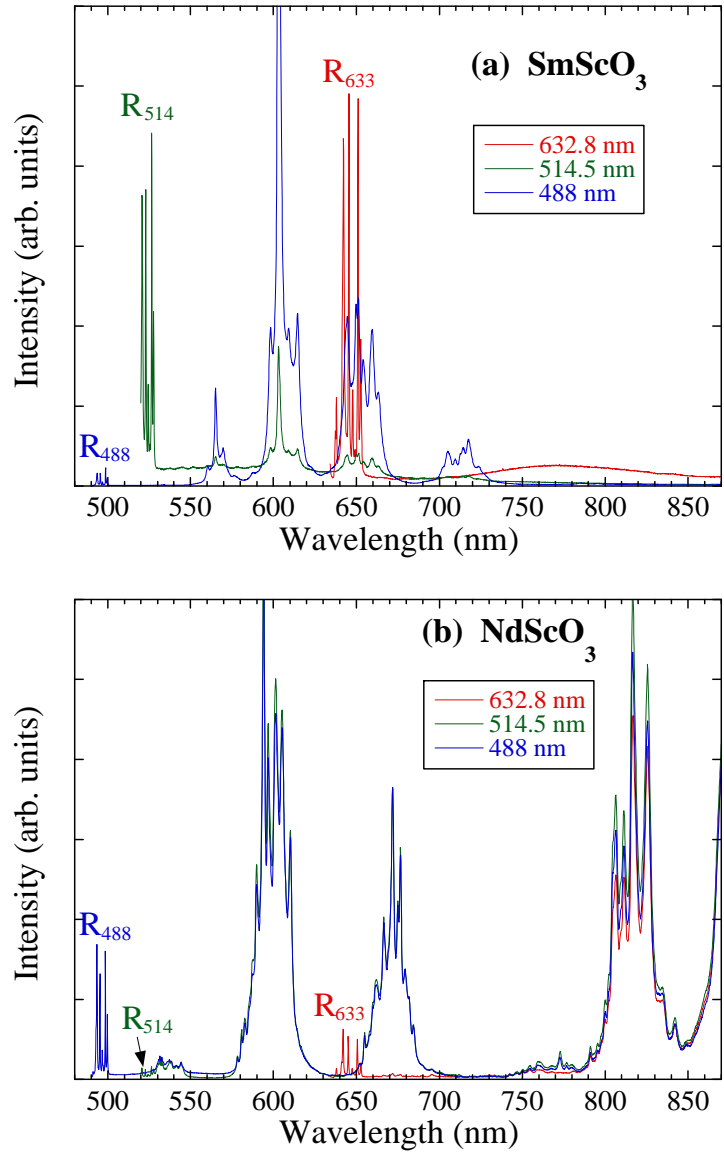

Figure 7